\documentclass[journal,twocolumn]{IEEEtran}

\usepackage{color}
\usepackage{graphicx}
\usepackage{amsfonts}
\usepackage{amsmath}
\usepackage{graphicx}
\usepackage{cite}
\usepackage{epsfig}
\usepackage{latexsym}
\usepackage{subfigure}
\usepackage{epstopdf}
\usepackage{color}
\usepackage{verbatim}
\usepackage{units}
\usepackage{amsthm}
\usepackage{hyperref}
\usepackage[shortlabels]{enumitem}

\begin{document}
\title{3GPP Network Architecture Enhancement for Ambient IoT Service}

\author{Dongjoo~Kim, Philippe~Godin, Bo~Bjerrum, Pallab~Gupta, \IEEEmembership{Member,~IEEE}, M.~Majid~Butt, ~\IEEEmembership{Senior~Member,~IEEE}

\thanks{Dongjoo Kim is with Nokia Standards, Seoul, Korea (e-mail: dongjoo.kim@nokia.com).
}
\thanks{Philippe Godin id with Nokia Standards, Massy, France (e-mail: philippe.godin@nokia.com).
}
\thanks{Pallab Gupta is with Nokia Standards, Bangalore, India (e-mail: pallab.gupta@nokia.com).
}

\thanks{Bo Bjerrum is with Nokia Standards, Aalborg, Denmark. (e-mail: bo.bjerrum@nokia.com.)
}
\thanks{M. Majid Butt is with Nokia Standards, Naperville, USA (e-mail: majid.butt@nokia.com).
}
}

\maketitle
\begin{abstract}
Ambient internet of things (A-IoT) paradigm is under study in 3GPP with the intention to provide a sustainable solution for 
the IoT market without any need to replace the batteries and operate in harsh environments where it is difficult to 
replenish batteries. This article provides insight on 3rd Generation Partnership Project (3GPP) 
discussions in Release 18 and 19 with the focus on network architecture aspects. 3GPP has recently decided to start 
normative work in its Radio Access Network (RAN) Working Group (WG) and discussions are ongoing to start a work item in 
other WGs with more focus on architecture aspects. We explore and analyze various aspects of system design related to architecture requirements to support A-IoT service, different architecture options to consider, security and authentication mechanisms for A-IoT devices as well as key challenges for standardization of A-IoT service.

\end{abstract}

\section{Introduction}
IoT (Internet of Things) landscape is rapidly expanding and increasing the demand for advanced communication frameworks to 
support massive deployment of devices. Fifth-Generation (5G) systems have introduced an 
innovative paradigm capable of supporting highly reliable communication protocols and a dense ecosystem of interconnected 
devices. Emergence of Sixth-Generation (6G) technology is expected to meet the growing demand for smarter, more 
adaptive, and resource-efficient infrastructures, further expanding these possibilities.
Market projections indicate that IoT adoption will continue to rise across various sectors, including urban development, 
precision agriculture \cite{arun_ambientiotcommunications}, medical systems, consumer electronics and automated 
manufacturing. The global energy harvesting systems market is expected to grow at a Compound Annual Growth Rate 
(CAGR) of $9.8 \%$ from 2024-2029 \cite{report_AIoT2}.

As proliferation of devices accelerates, the demand for these systems to be developed based on 
energy-efficient design principles is also increasing. This has sparked interest in sustainable energy solutions that 
harvest energy from environmental (ambient) sources, such as solar power, light energy, kinetic energy, and radio waves. 
These methods significantly reduce reliance on conventional power systems, minimizing maintenance demands and 
environmental impact.

Research in communication system design for Radio Rrequency (RF) powered devices has come a long way over the last decade or so. 
Significant progress has been made in various areas of system design including (but not limited to):
\begin{enumerate}
  \item System design considering RF energy harvesting models, e.g., linear vs non-linear energy harvesting models and circuit considerations \cite{energyharvesting_Bruno}.
  \item Physical system design for wireless powered devices including modulation, coding and waveform design 
      \cite{AIoT_waveforma_beamforming, AIoT_modulation}.
  \item Beamforming and radio resource management mechanism design for devices with intermittently available energy 
      \cite{AIoT_resourceallocation}.
  \item Medium Access Control (MAC) protocol design for low energy devices \cite{AIoT_accesscontrol}.
\end{enumerate}
In spite of availability of literature on technology fundamentals, wireless powered devices have not been yet adopted in 
mainstream communication systems. However, recent discussions within 3rd Generation Partnership Project (3GPP) have 
shown a strategic shift towards integrating innovative energy harvesting devices, also called Ambient IoT (A-IoT) devices, 
into future network architectures. A-IoT devices address the lowest segment for the IoT market below Narrow Band IoT 
(NB-IoT) and Reduced Capabilities (RedCap) devices in terms of data rate requirements targeting a few hundreds of bits and 
power consumption in the order of few (hundred) micro watts (µW).

3GPP started discussions on A-IoT devices in its Release 18 and decided to study them further in Release 19. The use 
cases, device types, network topologies and scenarios were identified for operation of these devices, 
and further use cases, topologies and scenarios will be added in future releases \cite{AIoT_SA}. Some recent works have captured this 
progress in standardization, but available literature lacks information on 3GPP progress regarding A-IoT services and 
architecture enhancements to enable A-IoT services in 3GPP networks \cite{AIoT_Nokia}.

This paper focuses on a critical aspect of these developments. It explores the ongoing progress in defining structural 
frameworks, key functional modules, and operational principles that will enable seamless integration of A-IoT devices into 
a 3GPP network. By discussing architectural design considerations and potential obstacles, we examine the broader 
impact of these innovations and identify pathways for realizing their full potential in a rapidly changing technological 
landscape.

%%%%%%%%%%%%%%%%%%%%%%%%%%%%%%%%%%%%%%%%%%%%%%%%%%%%%%%%%%%%%%%%%%%%%%%%%%%%%%%%%%%
%\begin{figure*}
%\centering
%  	\includegraphics[width=4.5in]{./Comparison_dev}
%   \caption{Comparison of device capabilities for different device types defined in 3GPP.}
%	\label{fig:dev}
%\end{figure*}

\begin{figure*}
\centering
  	\includegraphics[width=4.5in]{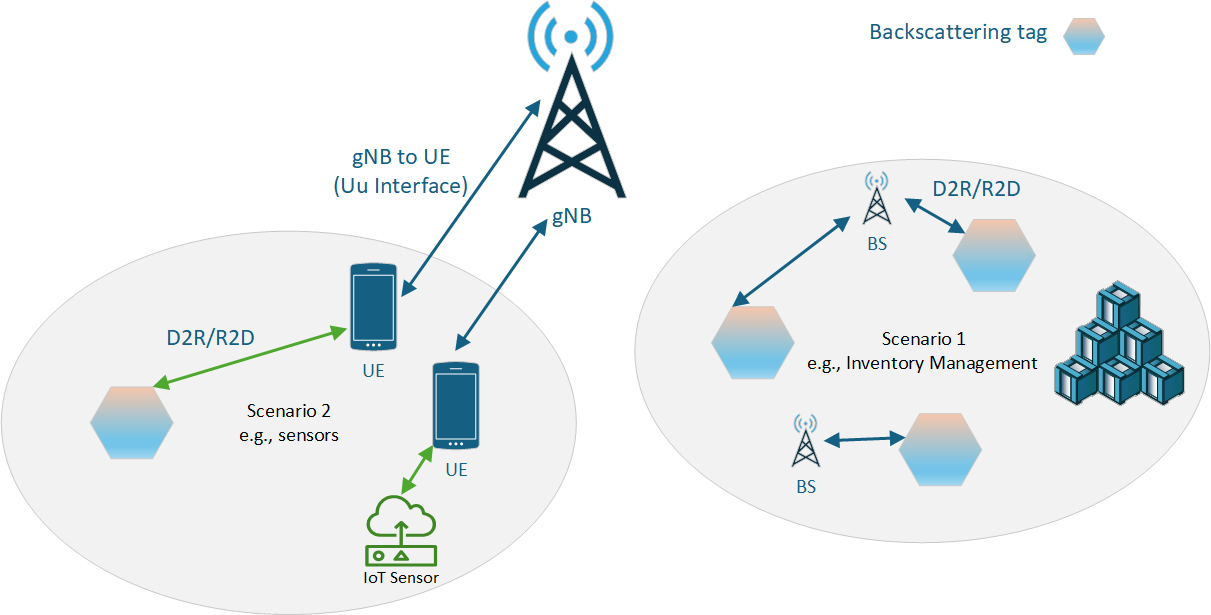}
   \caption{A-IoT scenarios and topologies as defined in 3GPP.}
	\label{fig:topologies}
\end{figure*}

\section{3GPP Work SCOPE AND TARGETS} 
\label{sect:devices}
In this section, we provide a summary of key features of A-IoT system based on device types, scenarios and communication topologies studied in 3GPP.
\subsection{A-IoT Device Types and Topologies in 3GPP} 
3GPP has defined 2 deployment scenarios for study in its Release 19 study item described in TR 38.848 \cite{AIoT_RANTR}. Scenario 1 
specifies indoor devices and indoor Base Stations (BS), while scenario 2 specifies indoor devices and outdoor base 
stations. A typical example of BS is a next generation Node B (gNB) as shown in scenario 2 in Fig.~\ref{fig:topologies}. 

To support the 2 scenarios, various device types and communication topologies have been defined as introduced 
below. 
\subsubsection{Device Types}
3GPP has defined 3 device types based on their capabilities. Two main technologies, backscattering and active internal transmission are specified. Device type 1 with backscattering has the lowest capabilities and device 2(b) with active transmission component has the highest capabilities.\\  
Device type 1 comprises the following capabilities:
\begin{enumerate}[(a)]
  \item Equipped with storage
\item Backscattering technology
\item 1 µW peak power consumption
\item	No uplink and donlink (UL/DL) amplifying
\end{enumerate}
Device type 2(a) has the following capabilities:
\begin{enumerate}[(a)]
  \item Equipped with storage
\item Backscattering technology
\item A few hundred µW peak power consumption
\item Both UL/DL amplifying
\end{enumerate}
Device type 2(b) includes the following capabilities:
\begin{enumerate}[(a)]
  \item Equipped with storage
\item UL transmission generated internally by the device
\item A few hundred µW peak power consumption
\item Both UL/DL amplifying
\end{enumerate}

\subsubsection{Communication Topologies}
The A-IoT system primarily consists of A-IoT devices capable of transmitting small amount of data (e.g. by backscattering) 
and A-IoT readers that can capture and process the transmitted data from the A-IoT devices. Depending on where the A-IoT 
reader functionality is hosted, there are two communication topologies defined:
\begin{itemize}
  \item \textbf{Topology 1} specifies direct BS to device communication and BS performs the reader functionality. This is 
feasible for indoor scenario when devices are close to indoor BS (scenario 1).
  \item \textbf{Topology 2} specifies communication through an intermediate device, such as user equipment (UE) as shown in 
Fig.~\ref{fig:topologies}, where the UE performs the reader functionality. This topology is applicable to both scenario 1 and 2, 
eliminating the requirement for BS to be in close proximity to the devices. The intermediate nodes could be network 
devices or UEs.
\end{itemize}

Fig.~\ref{fig:topologies} illustrates both scenarios and topologies, highlighting BS and UE as readers for topology 1 and 
2, respectively. Notably, the same access network nodes are shown to perform Device-to-Reader (D2R) and Reader-to-Device 
(R2D) communication. However, it is permitted for separate access network nodes to handle D2R and R2D functionalities in 
either topology.  

3GPP has concluded its study item and agreed to specify scenario 1, topology 1 and device type 1 during the normative 
phase in Release 19 \cite{AIOTRAN_WI19}. Scenario 2, topology 2 and other device types are planned for specification in Release 20, beginning 
in the second half of 2025.
\label{sect:arch}
\begin{figure*}
\centering
  	\includegraphics[width=5.5in]{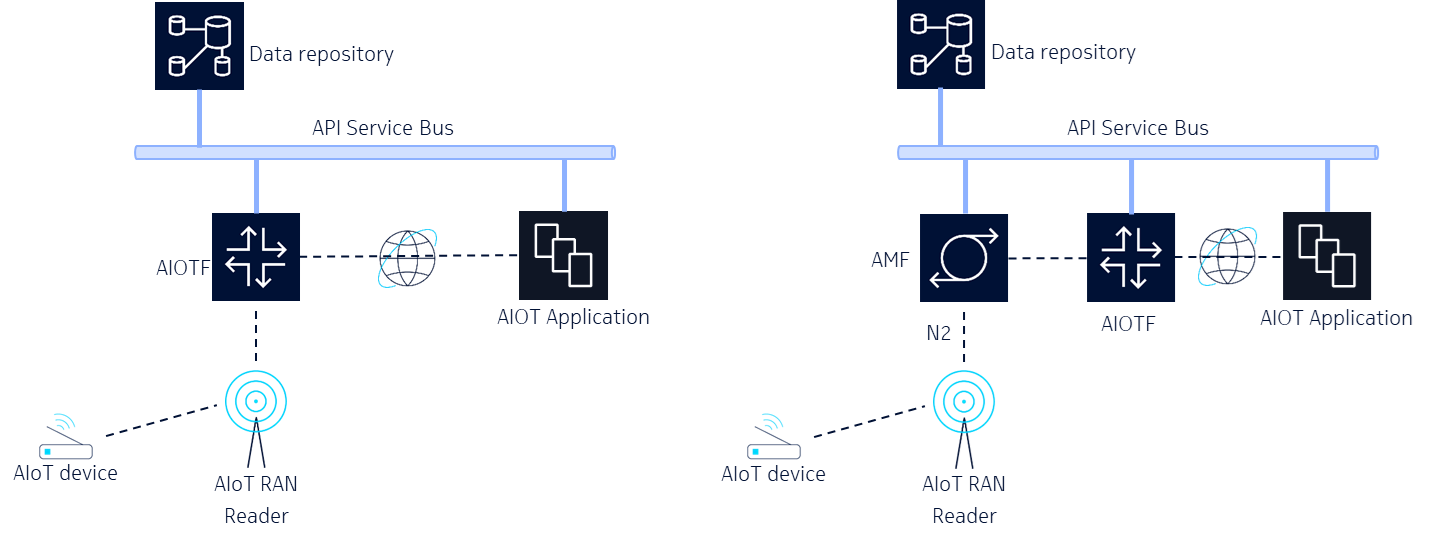}
   \caption{Architecture options to support Topology 1.}
	\label{fig:arch top1}
\end{figure*}

\subsection{A-IoT Use cases in 3GPP} 
 This section introduces some of the use cases and services that can be supported using A-IoT devices.
\begin{itemize}
  \item \textbf{Indoor inventory service} involves inventorying items in warehouses using A-IoT devices, enabling 
      efficient storage management through interaction with readers. The operations include verification, unloading, 
      gate-in/out, and inventory checks, as detailed in 3GPP TR 22.840 \cite{AIoT_SA}. Other use cases include medical 
      inventory in hospitals, intra-logistics in factories, and visitor guidance in museums.
  \item \textbf{Indoor sensors}, enabled by A-IoT devices, offer services like monitoring substations for safety, 
      ensuring proper storage of medical instruments, tracking environmental conditions for production, and deciding 
      smart laundry modes based on temperature and humidity.
  \item \textbf{Indoor command} for A-IoT devices include read, write, and activate/deactivate functions. The read 
      command queries sensor values (e.g., temperature, humidity) and reports them. The write command stores data in 
      devices with non-volatile memory, and the activate/deactivate command remotely manages device activation.
  \item \textbf{Indoor positioning} is supported at the reader granularity, meaning the device's location is determined 
      by the location of the reader to which it responds for services such as inventory and commands.
\end{itemize}
%%%%%%%%%%%%%%%%%%%%%%%%%%%%%%%%%%%%%%%%%%%%%%%%%%%%%%%%%%%%%%%%%%%%%%%%%%%%%%%%%%%%%%%%%%%%%%%%%%%%%%%%%%%%%%%%%%%%%%%%%%%%%%%%%%%%%%%%%%%%%%%

\section{Network ARCHITECTURE ASPECTS for A-IoT}

\subsection{General Architecture Principles}
3GPP has studied architecture enhancements needed to support A-IoT devices \cite{AIoT_SA2}. Some of the key principles 
that were considered for the architecture are:
\subsubsection{Traffic Pattern}
The architecture and procedure will support following types of traffic with A-IoT devices:
\begin{itemize}
  \item \textbf{Device Terminated (DT):} This type of messaging will be used by the network or an application (via the 
      network) to send a command or an inventory request to the A-IoT device.
\item	\textbf{Device Originated–Device Terminated Triggered (DO-DTT):} This type of messaging will be used by the 
    A-IoT device to respond back to a received inventory request or a command.
\item	\textbf{Device Autonomous (DO-A):} This type of messaging will be used by the A-IoT device to initiate communication session. This traffic type will not be specified in Release 19, but may be supported in later releases.
\end{itemize}

\subsubsection{Registration, Mobility and Connectivity states}
Since an A-IoT device is assumed to have extremely low complexity in design, a very small form factor, and is intended for 
specific use cases (e.g., warehouse automation, inventory management, etc.), the underlying assumption is that the conventional registration procedure used by general-purpose UEs is not required. Additionally, there is no requirement to 
support mobility or handover of A-IoT devices, and the network does not need to maintain Radio Resource Control (RRC) states of the devices.

\subsubsection{Deployment options}
For many of the use cases, data security and privacy are of utmost importance, and transmitting A-IoT data over the 
public network may not always be desirable feature for the A-IoT service provider. While the 5G network is extremely secure by 
design, to enhance data privacy and isolation, the 3GPP study also explored the possibility of deploying a dedicated 
network for A-IoT services.

\begin{figure*}
\centering
  	\includegraphics[width=5.5in]{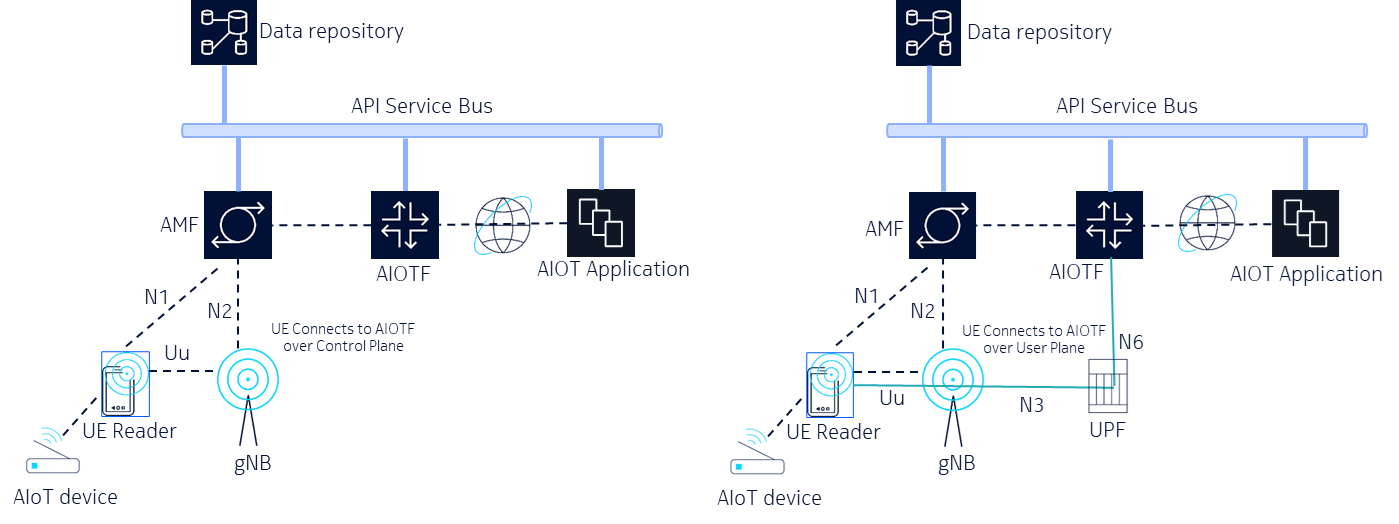}
   \caption{Architecture options to support Topology 2.}
	\label{fig:arch top2}
\end{figure*}

\subsection{Architecture Solutions} 
New functionalities are needed in the 5G Core (5GC) Network, for controlling the A-IoT device readers, securely exposing the 
A-IoT services capability of the network to applications, authorizing service requests from applications and validating and 
consolidating responses from A-IoT devices. To support these new functionalities a new core network function, called 
AIOTF, will be introduced in the 5GC. The AIOTF produces service-based Application Programming Interfaces (APIs) that can be consumed by, e.g., an A-IoT 
Application Function (AF) to access the AIoT services of the network. The AIOTF also uses the service based architecture to 
consume services produced by other network functions within 5GC.

Depending on the communication topology used for A-IoT deployments (i.e. topology 1 or topology 2), the AIOTF will need to 
support different ways to communicate with the A-IoT reader. The AIOTF connectivity will also vary depending on various 
deployment options. 

The following architecture options will be specified by 3GPP to cater different communication 
topologies and connectivity deployment options:

\subsubsection{Architecture For Topology 1}
In topology 1, the reader functionality is hosted in a Radio Access Network (RAN) node and referred as A-IoT RAN Reader. The A-IoT RAN Reader is embedded in the RAN node or deployed as a 
remote element managed by the A-IoT RAN node, such as a deported transmission point with the capability to perform A-IoT Device Reader functionality. As shown in Fig.~\ref{fig:arch top1}, there are two architecture options depending on how the AIOTF communicates with the 
A-IoT RAN Reader: in the first architecture option, the AIOTF communicates directly with the A-IoT RAN node whereas in the other option, the AIOTF communicates with the A-IoT RAN node indirectly via Access and Mobility Management Function (AMF).
 
The architecture option with direct communication between the AIOTF and the A-IoT RAN node without involving much of the 
5GC Network functions is suitable for some local and isolated deployments and requires a complete greenfield 
installation. On the other hand, the architecture with indirect communication between the AIOTF and the A-IoT RAN node is 
essential for brownfield deployments where an operator wants to re-use its existing 5G BS deployments for 
providing A-IoT services. This option is also suitable for deployment scenarios where a mix of communication topologies (i.e. 
topology 1 and 2) will be used.

\subsubsection{Architecture For Topology 2}
In Topology 2, the reader functionality is hosted in a UE. There are 
two architecture options possible depending on how the AIOTF communicates with the UE Reader as illustrated in 
Fig.~\ref{fig:arch top2}. In the first option, the AIOTF communicates with the UE Reader over the Control Plane signaling, 
whereas in the second alternative, the AIOTF communicates with the UE Reader over a User Plane connectivity established by 
the UE. In both architecture options, the base station serving the UE Reader is a gNB and this gNB  is in control of 
managing the radio resources used by the UE Reader to communicate with the A-IoT devices.

\begin{figure*}
\centering
  	\includegraphics[width=6.5in]{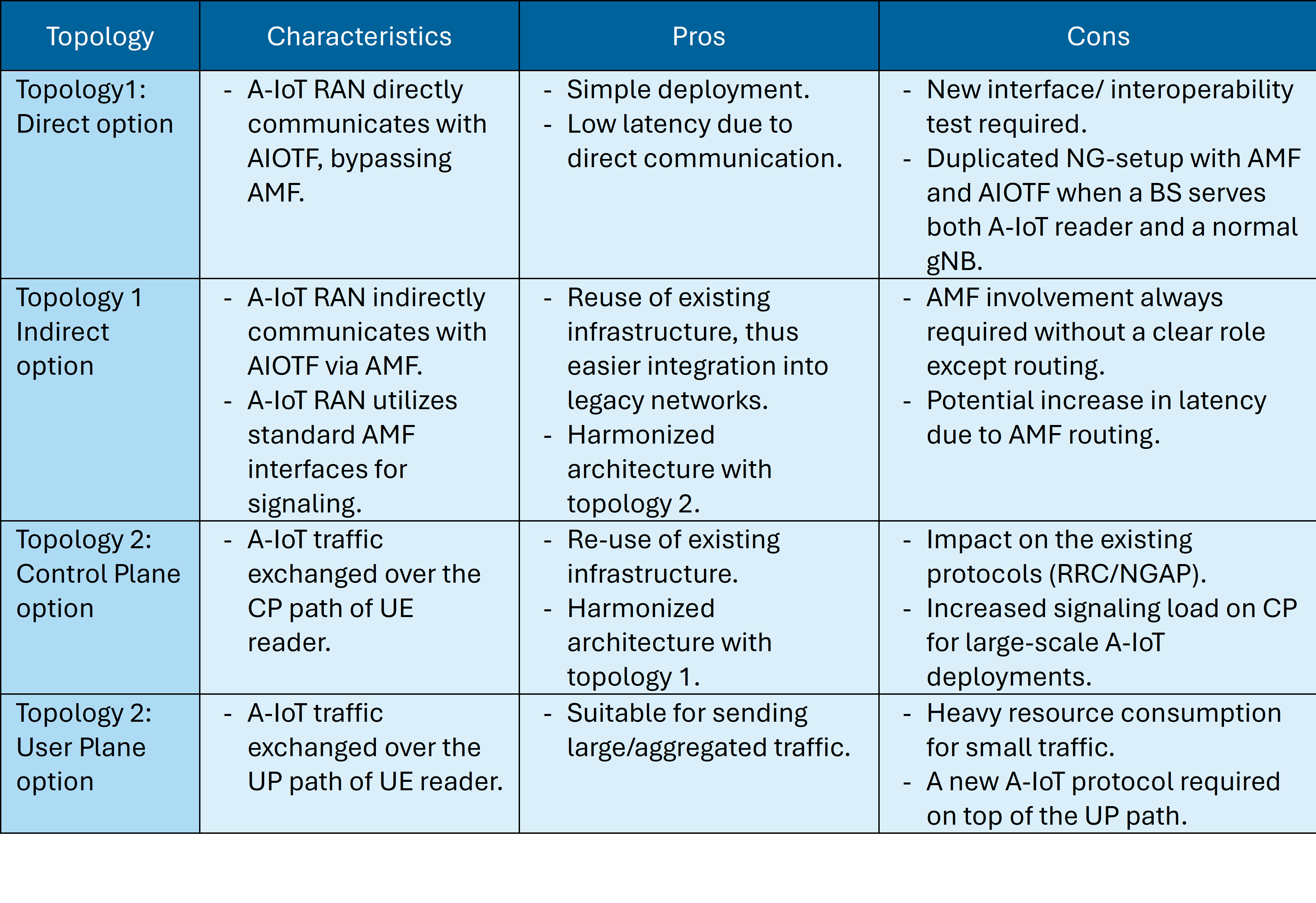}
   \caption{A comparison of architecture options for topology 1 and 2.}
	\label{fig:arch comp}
\end{figure*}

A comparison of pros and cons of architecture options for both topology 1 and 2 is presented in Fig.~\ref{fig:arch comp}. When both topologies are used simultaneously, the indirect option for topology 1 and the control plane option for topology 2 are preferable, as they can be supported by a unified architecture. In contrast, the direct option is better suited for standalone deployments, while the user plane option for topology 2 may well be more suitable for applications requiring support for large traffic volumes. 

\subsection{Protocol Stack}
Since A-IoT device’s hardware capability is highly limited and cannot support the existing Access Stratum (AS) layer 
protocol designed for NR UEs, a new A-IoT-specific AS protocol will be designed to facilitate communication between the 
A-IoT device and the A-IoT reader. A-IoT information exchanged over the A-IoT AS layer will be further communicated to the 
5GC via the A-IoT RAN node using the existing Next Generation Application Protocol (NGAP) protocol in topology 1, and 
via the UE reader and the gNB using RRC-based Uu protocols and the NGAP protocol, respectively, in topology 2. For 
topology 2, in the alternative where the AIOTF and the UE Reader communicates over a User Plane connectivity (PDU Session) 
established by the UE, a different protocol stack will be used between the UE reader and the AIOTF. 
Further, a new Non-Access Stratum (NAS) layer protocol will be defined to support communication between the A-IoT device 
and the AIOTF. This A-IoT-specific NAS layer will handle A-IoT service requests (e.g., inventory) and responses from the 
A-IoT device. The simplified protocol stack is illustrated in Fig.~\ref{fig:protocol_stack} to provide an overview.

\section{DEVICE IDENTIFICATION AND SUBSCRIPTION}

\subsection{A-IoT Device Identifier}
Each A-IoT device is assigned a unique identifier, which may be provided by the operator or a third-party enterprise, 
depending on the ownership. The identifier consists of two parts: the first part indicates the ID type, including the 
A-IoT network identifier and/or the third-party enterprise identifier, while the second part contains a device ID, similar 
to Electronic Product Code (EPC) that uniquely identifies the device \cite{GS1}. The combination of both parts 
ensures global uniqueness.
\subsection{A-IoT Device Information Management}
The number of A-IoT devices can be significantly greater than the number of general-purpose UEs (e.g. smart phones) served 
by a network. Additionally, A-IoT devices consume fewer network operator provided services than conventional 
UEs. For these reasons, A-IoT devices are not expected to have UE-like subscription information within the network 
operator. However, when an A-IoT device is managed by the network, some information may be stored by the network 
operator to allow the A-IoT devices to make a better use of network resources. If A-IoT Devices are not managed by the 
network operator, such A-IoT device information for authentication and/or authorization may be stored in an external Authentication, Authorization and Accounting (AAA) 
server. In addition to this static information about A-IoT devices, dynamic device information such as the device 
authorization context, the last-used A-IoT Reader, etc., may also be stored to optimize system procedures. 

\section{PROCEDURE ASPECTS of AIoT Service}

\subsection{A-IoT Device Registration and Provisioning}
To effectively manage A-IoT devices, network operators must obtain essential information, such as device identification 
and security credentials through 3GPP networks. However, due to hardware limitations, A-IoT devices cannot accommodate 
Universal Integrated Circuit Cards (UICCs) or implement the conventional registration procedures designed for general-purpose UEs. To address this, 3GPP has 
conducted research on alternative mechanisms. One proposed approach involves third parties statically provisioning device 
information before any interaction between A-IoT devices and the network. While this approach enables the network to 
acquire information about A-IoT devices, it has the drawback of limiting the flexibility of use cases. To resolve these 
issues, further studies are required to explore methods for obtaining A-IoT device information in a more dynamic and 
adaptable manner.
\subsection{A-IoT Reader Authorization}
It is essential for an A-IoT reader to be authorized by a 3GPP network before functioning in its designated role. In 
topology 1, where the A-IoT RAN reader is deployed in a pre-planned manner by the network operator, no standardized 
authorization mechanism is deemed necessary due to the controlled deployment process. Conversely, in topology 2, the UE 
reader requires network authorization. To address this, the subscription data of a UE will be enhanced to include an 
indicator specifying its authorization to function as an A-IoT reader. The UE's authorization is then validated during the 
registration procedure using this enhanced subscription data and may be delivered to the gNB. Further studies are planned 
to determine whether the UE reader requires additional authorization steps when establishing a PDU session towards the 
AIOTF if the UP option of topology 2 is used. This investigation aims to ensure that only UEs authorized for the UP option are permitted to establish a PDU session to connect to the AIOTF.

\subsection{Information Provided by AF}
3GPP network design ensures that the A-IoT services are always initiated by an AF. When the AF 
requests A-IoT services from the AIOTF via the Network Exposure Function (NEF), it provides necessary information for 
the 5G system to execute the required procedures. The AF specifies the A-IoT service type, such as inventory or command, 
indicating the requested service. Additionally, it may provide information that enables the AIOTF to select appropriate 
A-IoT readers, such as geographical location, pre-defined area information, or one or more particular targeted UE reader IDs, when 
applicable. The AF may also provide information about the target A-IoT devices and their capabilities that helps the BS to allocate resources for communication between the A-IoT reader and A-IoT devices.
\subsection{AIOTF Selection}
When the NEF receives an A-IoT service request from the AF, it selects an appropriate AIOTF to perform the requested 
operation. While the specific details are still under discussion within 3GPP, one proposed method involves the AIOTF 
registering with the Network Repository Function (NRF) and providing its NF profile along with its service area information. In this approach, the NEF 
can use the area information included in the request received from the AF to identify a suitable AIOTF through the NRF. If 
the AF provides a specific UE reader ID instead of area information, the NEF must first determine the location of the 
given UE reader before identifying an appropriate AIOTF.  
\subsection{A-IoT Reader Selection}
3GPP has agreed on some general principles for selecting A-IoT readers. 
\begin{itemize}
  \item First, the AIOTF is responsible for selecting a 
list of suitable A-IoT readers. In topology 1, this list refers to A-IoT RAN readers, while in topology 2, it refers to 
authorized UE readers.
  \item Second, if the A-IoT service operator has pre-installed A-IoT readers and already knows their 
location information (static deployment option), the AIOTF can map the A-IoT readers based on location information or 
pre-agreed area ID information provided by the AF.
  \item Third, if the A-IoT service operator is unaware of the deployed A-IoT 
readers (dynamic deployment option, applicable only to topology 2), the AIOTF can generate a candidate list of UE readers 
based on essential network information such as location, mobility, and authorization status. This list is provided to the 
gNB, which can further refine it based on radio conditions, such as signal strength, A-IoT device capabilities, and 
frequency etc. 
\end{itemize}

\begin{figure}[t!]
\centering
  	\includegraphics[width=3.5in]{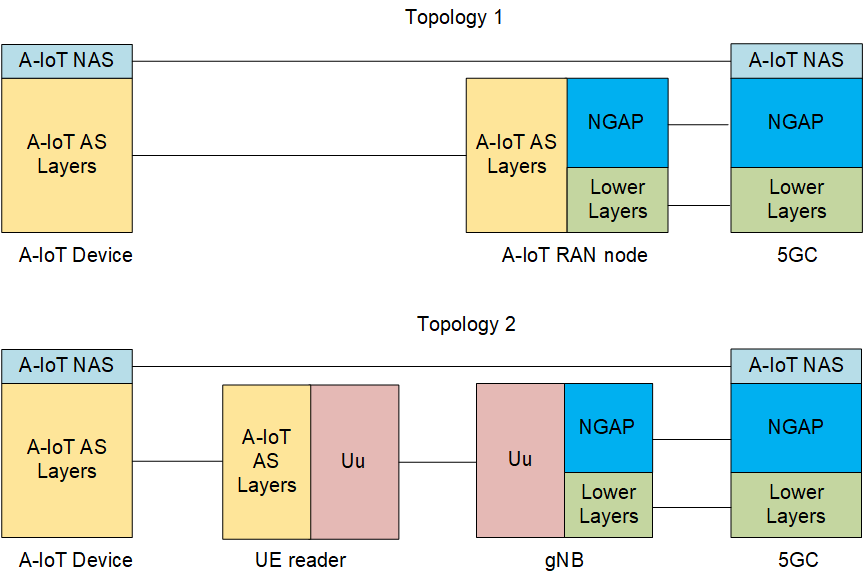}
   \caption{Simplified protocol stacks for topology 1 and topology 2.}
	\label{fig:protocol_stack}
\end{figure} 
\subsection{Radio Resource Allocation}
Since the 3GPP A-IoT system operates over a licensed frequency band, usage of radio resources must be controlled by 
the A-IoT RAN.  To support the A-IoT RAN to manage the radio resources, the AIOTF may provide some assistance information 
to the A-IoT RAN, that includes information like the A-IoT service type (e.g., inventory), the approximate 
D2R response size, and the estimated number of A-IoT devices based on the AF request. Additional 
assistance information may be introduced as the work progresses such as Session Identifier, A-IoT device capability, 
estimated number of responses \cite{AIoT_RANTR19}. In future studies, further details of the information as well as mechanisms to provide 
it to the A-IoT RAN will be explored.
\subsection{Overall Message Flow} 

\begin{figure*}
\centering
  	\includegraphics[width=4.5in]{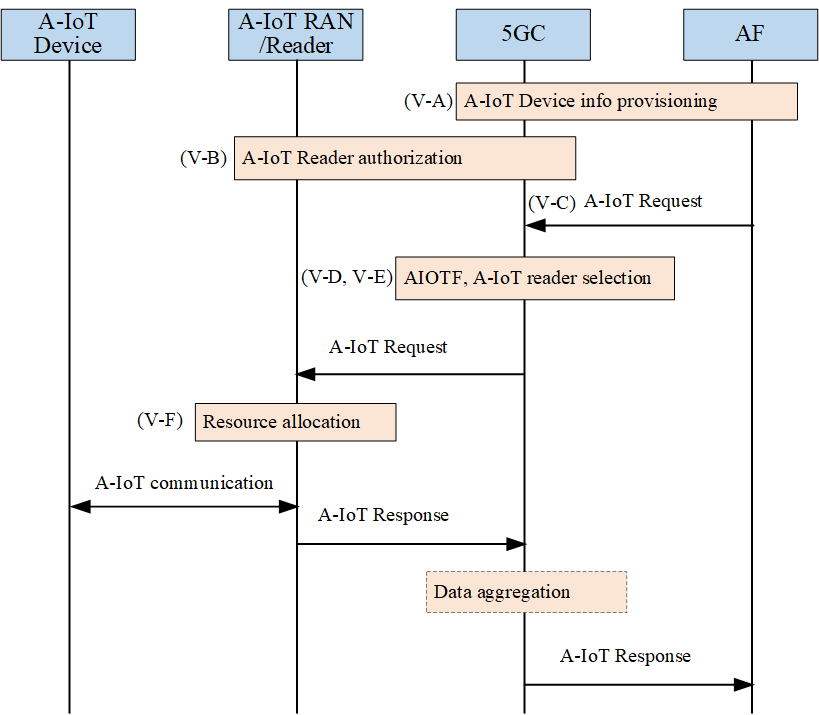}
   \caption{Overall service flow for A-IoT service operations.}
	\label{fig:service_flow}
\end{figure*} 
Fig.~\ref{fig:service_flow} provides an overview of the A-IoT service flow. To begin with, A-IoT device information is 
provisioned from the AF to the 5GC, and A-IoT Readers are authorized. The AF sends an A-IoT request to the 5GC, which 
authorizes the request and selects the appropriate AIOTF and BS. The request, along with the A-IoT task ID, is 
forwarded to the BS, which allocates radio resources. The Reader collects the responses, including the A-IoT 
device identifier, task ID, and additional data, which are then aggregated by the 5GC and sent back to the AF.
\section{SECURITY ASPECTS for A-IoT Service} 
\label{sec:security}
The A-IoT device is a fundamentally different type of device compared to traditional UE (e.g. smartphone) and will not follow UE’s traditional registration and communicating mechanism through 5G network. For example, the trust model of the UE relies on the existence of a UICC holding the credentials used to authenticate it to the network. As the A-IoT device is both constrained in size and power consumption, neither UICC based authentication nor classical authentication methods will be feasible. This calls for new trust models and procedures. 3GPP SA3 WG has studied aspects of privacy, authentication, authorization, integrity, confidentiality and anti-replay, all to ensure similar security level as provided by existing 3GPP features \cite{AIoT_SA3}. As the capabilities of the A-IoT devices will vary based on their constraints and the deployment scenario, there is a collective understanding in 3GPP SA3 to provide the means for reaching the same security level as existing 3GPP features. Evaluation of which security capabilities are required in a particular deployment and use case is left to the network operators. 

The A-IoT service authentication, integrity, confidentiality and anti-replay primitives and procedures are quite different compared to classical methods as the signaling between A-IoT device and the network is limited to a few messages, wherein multiple primitives, depending on the required capabilities, are needed to be executed and terminated to establish a secure foundation for exchanging the identity or data. These new methods extend 3GPP technology into encompassing a new set of devices, aimed for low power IoT and open new usages patterns for the technology.

The protocol stack shown in Fig.~\ref{fig:protocol_stack} gives a simplified view of the layers for the given topologies. A common denominator of the stacks is the AIoT NAS layer, which is the end-to-end connection between 5GC and AIoT device. Establishing security at the AIoT NAS layer makes security procedures agnostic to topology and reduces the trust base to endpoints only, in contrast to hop-by-hop based security approaches. Topology 2 with a UE Reader is even more challenging as UE in general is an untrusted entity and it is not clear how to ensure that a UE cannot impersonate a privileged UE. The answer lies within the reuse of the primary authentication of the reader UE. The UICC based authentication provides evidence of identity of the user and hereby, the privilege to act as a reader, while similar challenges do not exist for topology 1 as the gNB is part of the CN trust domain.

The deployment of an A-IoT system configuration requires careful evaluation of the use case involved. As the framework only provides the means to secure communication between the AIOTF and AIoT device, it is left to the configuration to define which means will be enabled. For example, if the AIoT device is attached to a high value asset in a warehouse, it will be of interest of eavesdroppers to know the quantity or impersonate the asset and therefore authentication, anti-replay and privacy means needs to be enabled. On the flip side, it does not make sense to enable the same protection level for low value assets. Finding acceptable trade-off between AIoT device complexity and required protection means will impose a risk to the deployments. It is recommended to assess each use case and carefully evaluate required protection means.  

\section{FUTURE CHALLENGES}  
The future challenges for the 3GPP A-IoT system are multifaceted and demand innovative solutions. Some of 
these challenges are summarized below:
\begin{enumerate}
  \item Scalability will be a critical concern as the system must support a large number of devices while maintaining 
      network performance and resource efficiency. 
  \item Ensuring inter-operability across diverse A-IoT device types with varying capabilities and different topologies 
      will require robust standardization and seamless integration mechanisms. 
  \item A-IoT devices have no RRC states. Legacy paging and access functionalities have been designed based on the RRC states 
      and corresponding transition mechanisms. System design without RRC states requires a new design philosophy and 
      corresponding changes over air interface as well as core network.
  \item Security and privacy will remain of paramount importance, as the increased exchange of sensitive data necessitates scalable and lightweight protection measures.   
  \item Energy efficiency poses a significant challenge, particularly for devices with constrained power sources, making 
      it essential to develop mechanisms that minimize energy consumption. The system must also adapt dynamically to 
      varying service demands, such as real-time monitoring and long-term data collection, while maintaining 
      high-quality service levels.
  \item Connection management and tracking of A-IoT devices is also critical, e.g. use cases where the A-IoT devices are used in shipments. Since the A-IoT devices do not register with the network and no RRC states are 
      maintained, it requires a completely new mechanism for device tracking and connection management including re-designing of the roaming architecture to support A-IoT devices and services.
  %\item Some of the outdoor use cases, e.g., tracking of shipments and containers equipped with A-IoT devices may require A-IoT service continuity with change of Mobile Network Operator (MNO). 
     
  \item Solutions to enhance the accuracy of locating AIoT device need to be worked out. The current 3GPP agreement of simple derivation of the AIoT device location from the location of an associated reader may not suffice to meet the requirements of some of the indoor positioning use cases identified in TR 38.848 \cite{AIoT_RANTR}, where the location of the AIoT device must be determined within an accuracy of 1-3 meters with $90 \%$ confidence.  
  \item Finally, balancing deployment and maintenance costs with affordability for end-users and stakeholders will be a persistent issue.
\end{enumerate}

\section{Conclusion}
\label{sect:conclusions}
This article provides a timely update on the status of A-IoT design in 3GPP, with a focus on architecture, deployment 
options, and key challenges. 
%A-IoT devices, designed to operate at ultra-low power consumption and harness energy from 
%ambient sources, represent a significant step toward sustainable IoT solutions. 
We examine and compare various device types, communication 
topologies, and deployment scenarios outlined in 3GPP Release 19 and discuss their implications for system design. The 
RAN WGs in 3GPP have agreed to start a work item in Release 19, while the SA WGs are expected to follow and 
start with normative phase in Release 19 after addressing security concerns for the A-IoT system. Security is a key concern 
for A-IoT devices as it is challenging to implement legacy security protocols due to their low complexity and cost. Unlike Radio Frequency Identification (RFID) devices, 3GPP has more stringent requirements on security for devices and it is important to agree on low complexity 
security protocols before standardizing them. Finding the optimum security configuration for a given use case will be a case-by-case evaluation, and as the technology matures, further protection can be applied for the security-constrained devices. 3GPP standardization of the A-IoT system offers significant potential to advance sustainable IoT networks and overcoming challenges in their operation, scalability, security, and energy efficiency will be essential to realize A-IoT service with full potential.

\bibliographystyle{IEEEtran}
\bibliography{bibliography}

% Generated by IEEEtran.bst, version: 1.14 (2015/08/26)
\begin{thebibliography}{10}
\providecommand{\url}[1]{#1}
\csname url@samestyle\endcsname
\providecommand{\newblock}{\relax}
\providecommand{\bibinfo}[2]{#2}
\providecommand{\BIBentrySTDinterwordspacing}{\spaceskip=0pt\relax}
\providecommand{\BIBentryALTinterwordstretchfactor}{4}
\providecommand{\BIBentryALTinterwordspacing}{\spaceskip=\fontdimen2\font plus
\BIBentryALTinterwordstretchfactor\fontdimen3\font minus
  \fontdimen4\font\relax}
\providecommand{\BIBforeignlanguage}[2]{{%
\expandafter\ifx\csname l@#1\endcsname\relax
\typeout{** WARNING: IEEEtran.bst: No hyphenation pattern has been}%
\typeout{** loaded for the language `#1'. Using the pattern for}%
\typeout{** the default language instead.}%
\else
\language=\csname l@#1\endcsname
\fi
#2}}
\providecommand{\BIBdecl}{\relax}
\BIBdecl

\bibitem{arun_ambientiotcommunications}
\BIBentryALTinterwordspacing
A.~N. Arun, B.~Lee, F.~A. Castiblanco, D.~R. Buckmaster, C.-C. Wang, D.~J.
  Love, J.~V. Krogmeier, M.~M. Butt, and A.~Ghosh, ``Ambient {IoT}:
  Communications enabling precision agriculture,'' 2024. [Online]. Available:
  \url{https://arxiv.org/abs/2409.12281}
\BIBentrySTDinterwordspacing

\bibitem{report_AIoT2}
\BIBentryALTinterwordspacing
``Energy harvesting systems market size \& share analysis - growth trends \&
  forecasts (2025 - 2030),'' Mordor Intelligence, India, Tech. Rep., 2024.
  [Online]. Available:
  \url{https://www.mordorintelligence.com/industry-reports/energy-harvesting-system-market}
\BIBentrySTDinterwordspacing

\bibitem{energyharvesting_Bruno}
B.~Clerckx, R.~Zhang, R.~Schober, D.~W.~K. Ng, D.~I. Kim, and H.~V. Poor,
  ``Fundamentals of wireless information and power transfer: From {RF} energy
  harvester models to signal and system designs,'' \emph{IEEE Journal on
  Selected Areas in Communications}, vol.~37, no.~1, pp. 4--33, 2019.

\bibitem{AIoT_waveforma_beamforming}
S.~Shen and B.~Clerckx, ``Joint waveform and beamforming optimization for
  {MIMO} wireless power transfer,'' \emph{IEEE Transactions on Communications},
  vol.~69, no.~8, pp. 5441--5455, 2021.

\bibitem{AIoT_modulation}
M.~Varasteh, J.~Hoydis, and B.~Clerckx, ``Learning to communicate and energize:
  Modulation, coding, and multiple access designs for wireless
  information-power transmission,'' \emph{IEEE Transactions on Communications},
  vol.~68, no.~11, pp. 6822--6839, 2020.

\bibitem{AIoT_resourceallocation}
Z.~Wei, X.~Yu, D.~W.~K. Ng, and R.~Schober, ``Resource allocation for
  simultaneous wireless information and power transfer systems: A tutorial
  overview,'' \emph{Proceedings of the IEEE}, vol. 110, no.~1, pp. 127--149,
  2022.

\bibitem{AIoT_accesscontrol}
L.~Zhang, G.~Feng, S.~Qin, Y.~Sun, and B.~Cao, ``Access control for ambient
  backscatter enhanced wireless internet of things,'' \emph{IEEE Transactions
  on Wireless Communications}, vol.~21, no.~7, pp. 5614--5628, 2022.

\bibitem{AIoT_SA}
\BIBentryALTinterwordspacing
3rd Generation Partnership Project~(3GPP){, TR 22.840}, ``Study on ambient
  power-enabled internet of things,'' Tech. Rep., 2023. [Online]. Available:
  \url{https://www.3gpp.org/ftp/Specs/archive/22_series/22.840/22840-120.zip}
\BIBentrySTDinterwordspacing

\bibitem{AIoT_Nokia}
M.~M. Butt, N.~R. Mangalvedhe, N.~K. Pratas, J.~Harrebek, J.~Kimionis,
  M.~Tayyab, O.-E. Barbu, R.~Ratasuk, and B.~Vejlgaard, ``Ambient {IoT}: A
  missing link in {3GPP IoT} devices landscape,'' \emph{IEEE Internet of Things
  Magazine}, vol.~7, no.~2, pp. 85--92, 2024.

\bibitem{AIoT_RANTR}
\BIBentryALTinterwordspacing
3rd Generation Partnership Project~(3GPP){, TR 38.848}, ``Ambient {IoT}
  (internet of things) in {RAN},'' Tech. Rep., 2023. [Online]. Available:
  \url{https://portal.3gpp.org/desktopmodules/Specifications/SpecificationDetails.aspx?specificationId=4146}
\BIBentrySTDinterwordspacing

\bibitem{AIOTRAN_WI19}
\BIBentryALTinterwordspacing
3rd Generation Partnership Project~(3GPP){, RP-243326}, ``New work item:
  Solutions for ambient {IoT} (internet of things) in {NR},'' Tech. Rep., Dec.
  2024. [Online]. Available:
  \url{https://www.3gpp.org/ftp/Meetings_3GPP_Sync/RAN/Inbox/}
\BIBentrySTDinterwordspacing

\bibitem{AIoT_SA2}
\BIBentryALTinterwordspacing
3rd Generation Partnership Project~(3GPP){, TR 23.700-13}, ``Study on
  architecture support of ambient power-enabled internet of things (release
  19),'' Tech. Rep., Dec. 2024. [Online]. Available:
  \url{https://portal.3gpp.org/desktopmodules/Specifications/SpecificationDetails.aspx?specificationId=4255}
\BIBentrySTDinterwordspacing

\bibitem{GS1}
\BIBentryALTinterwordspacing
GS1, ``{EPC} tag data standard ({TDS}), release 2.1,'' Tech. Rep., Feb 2024.
  [Online]. Available: \url{https://www.gs1.org/standards/tds}
\BIBentrySTDinterwordspacing

\bibitem{AIoT_RANTR19}
\BIBentryALTinterwordspacing
3rd Generation Partnership Project~(3GPP){, TR 38.769}, ``Study on solutions
  for ambient {IoT} (internet of things) in {NR}.'' Tech. Rep., Dec. 2024.
  [Online]. Available:
  \url{https://portal.3gpp.org/desktopmodules/Specifications/SpecificationDetails.aspx?specificationId=4285}
\BIBentrySTDinterwordspacing

\bibitem{AIoT_SA3}
\BIBentryALTinterwordspacing
3rd Generation Partnership Project~(3GPP){, TR 33.713}, ``Study on security
  aspect of ambient {IoT} services in {5G} (release 19).'' Tech. Rep., Dec.
  2024. [Online]. Available:
  \url{https://portal.3gpp.org/desktopmodules/Specifications/SpecificationDetails.aspx?specificationId=4291}
\BIBentrySTDinterwordspacing

\end{thebibliography}
%\begin{IEEEbiography}[{\includegraphics[width=1in,height=1.25in,clip,keepaspectratio]{Majid.eps}}]{M. Majid Butt}(S'07 -- 
%M'10 -- SM'15) 
%\end{IEEEbiography}
%
%
%

%\end{IEEEbiography}

\end{document}